\documentclass[prd,showpacs,epsf,preprintnumbers
]{revtex4}

\renewcommand{\theequation}{\arabic{section}.\arabic{equation}}
\def\lb{\label}
\def\bb{\bibitem}
\def\be{\begin{equation}}
\def\ee{\end{equation}}
\def\ba{\begin{eqnarray}}
\def\ea{\end{eqnarray}}
\def\ds{\displaystyle}
\def\ol{\overline}
\def\rd{{\rm d}}
\def\rD{{\rm D}}
\def\e{{\rm e}}
\def\nn{\nonumber}
\def\5{_{(5)}}
\def\4{_{(4)}}
\def\2{\sqrt{2}}
\def\3{\sqrt{3}}

\def\X{{\bf X}}
\def\S{{\bf S}}
\def\R{{\cal R}}
\def\L{{\bf L}}
\def\a{{\bf \alpha}}
\def\b{{\bf \beta}}
\def\c{{\bf \gamma}}

\begin{document}
\begin{flushright}LAPTH-026/13
\end{flushright}
\begin{flushright}DTP-MSU/14-13
\end{flushright}
\title{
Bertotti-Robinson solutions of $D=5$
Einstein-Maxwell-Chern-Simons-Lambda theory  }
\author
{Adel Bouchareb}\email{adel.bouchareb@univ-annaba.org} \affiliation{
Institute of Physics, BM Annaba University, \\
 B.P. 11, 23000 Annaba, Algeria}
\author
{Chiang-Mei Chen}\email{cmchen@phy.ncu.edu.tw} \affiliation{
Department of Physics and Center for Mathematics and Theoretical
Physics,  \\
  National Central University, Chungli 320, Taiwan }
\author
{G\'erard Cl\'ement}\email{gerard.clement@lapth.cnrs.fr}
\affiliation{
LAPTh, Universit\'e de Savoie, CNRS, 9 chemin de Bellevue, \\
BP 110, F-74941 Annecy-le-Vieux cedex, France}
\author{Dmitri V. Gal'tsov} \email{galtsov@phys.msu.ru}
\affiliation{Department of Theoretical Physics, Moscow State
University, 119899, Moscow, Russia}

\begin{abstract}
We present a series of new solutions in five-dimensional
Einstein-Maxwell-Chern-Simons theory with an arbitrary Chern-Simons
coupling $\gamma$ and a cosmological constant $\Lambda$. For general
$\gamma$ and $\Lambda$ we give various generalizations of the
Bertotti-Robinson solutions supported by electric and magnetic
fluxes, some of which presumably describe the near-horizon regions
of black strings or black rings. Among them there is a solution
which could apply to the horizon of a topological AdS black ring in
gauged minimal supergravity. Others are horizonless and geodesically
complete. We also construct extremal asymptotically flat
multi-string solutions for $\Lambda=0$ and arbitrary $\gamma$.
\end{abstract}

\maketitle

\section{Introduction}

Five-dimensional supergravity is an interesting proving ground for
string theory. Its Lagrangian, obtainable by toroidal dimensional
reduction of eleven-dimensional supergravity, contains a
Chern-Simons term for the Maxwell field inherited from reduction of
the corresponding four-form term. Though it does not influence the
Einstein equations, it modifies the Maxwell equations and
consequently the gravitational field too. Surprisingly enough, its
presence is crucial for the enhancement of hidden symmetries of
five-dimensional Einstein-Maxwell theory in the case of field
configurations possessing two commuting spacetime Killing vectors.
For such configurations the theory reduces to a three-dimensional
sigma model realizing a harmonic map from the spacetime manifold to
the homogeneous space $G/H=G_{2(+2)}/((SL(2,R)\times SL(2,R))$
\cite{g2,5to3,BerkPiol,Gal'tsov:2008nz}. Owing to this hidden
symmetry, a generating technique  had been developed \cite{g2,5to3},
which opened the way to derive new charged rotating black rings and
general black strings
\cite{Gal'tsov:2009da,TYM,Compere09,Figu,KHPV,Compere10}.

In various physical contexts one is also interested in the more
general Einstein-Maxwell theory containing a Chern-Simons term with
an arbitrary coupling constant $\gamma$ and a cosmological constant
$\Lambda$:
 \be\lb{emcs5}
S_5 = \frac1{16\pi G_5}\int \rd^5x \bigg[\sqrt{|g_{(5)}|} \bigg(R\5
- \frac14F\5^{\mu\nu}F_{(5)\mu\nu}-2\Lambda \bigg)    -
\frac\gamma{12\sqrt3}\epsilon^{\mu\nu\rho\sigma\lambda}F_{(5)\mu\nu}
F_{(5)\rho\sigma}A_{(5)\lambda} \bigg]\,,
 \ee
where $F\5 = \rd A\5$, $\mu,\nu,\cdots = 1,\cdots,5$, and
$\epsilon^{\mu\nu\rho\sigma\lambda}$ is an antisymmetric symbol
whose signs will be detailed later. For $\gamma=1$, $\Lambda =0$
this is the action of  minimal five-dimensional supergravity, for
$\gamma=1$, $\Lambda <0$ the action of minimal gauged supergravity,
while for $\gamma=0$ it is the Einstein-Maxwell (EM) action.
Restricted to field configurations possessing two commuting Killing
vectors, this theory reduces to a three-dimensional gravitating
sigma model coupled to a potential originating from the cosmological
constant term. For $\gamma\neq 1$ the target space of this sigma
model is not a symmetric space (the isometry group is solvable), so
there are no non-trivial hidden symmetries which could be used to
generate exact solutions. Moreover, the potential term is not
invariant under target space isometries apart from some trivial
ones. Not surprisingly, no exact charged rotating black hole
solutions are known in the pure Einstein-Maxwell ($\gamma=0$) theory
even for $\Lambda=0$, though their existence was demonstrated
perturbatively \cite{AF,NL,AKN} and numerically \cite{KNP}. A
similar situation holds for charged black rings: static solutions in
EM theory have been found by Ida and Uchida \cite{IU} , and
generalized to EM-dilaton theory in \cite{KL,Y}. But {\em stationary
charged} black ring solutions of pure EM theory are not known in a
closed form, though their existence again was confirmed both
perturbatively solutions (for small charges \cite{OP}) and
numerically \cite{Kleihaus:2010hd}. For $\gamma\neq 1,0$ exact
charged black hole/ring solutions are not known either, while
numerical black hole solutions have been constructed
\cite{Kunz:2006xk} and shown to have unusual properties for $\gamma
> 1$ such as rotation in the sense opposite to the angular
momentum and a negative horizon mass with positive asymptotic mass.

It is therefore of interest to explore other tools to generate exact
solutions of the general EMCSL action (\ref{emcs5}) which could shed
some light on the nature of black objects in this theory. A
particular motivation for this  lies in the still unsolved question
about the existence of asymptotically $AdS$ and $dS$ black rings in
presence of the cosmological constant. The issue of charged
five-dimensional black objects with cosmological constant was
discussed previously in a number of papers. Supersymmetric $AdS_5$
black holes were obtained by several authors \cite{GR}. General
non-extremal rotating black holes in minimal five-dimensional gauged
supergravity were constructed by Chong, Cveti\u{c}, L\"u and Pope
\cite{CCLP}. Charged squashed black holes in EM theory ($\gamma=0$)
with cosmological constant were constructed numerically in
\cite{BRS}. Black strings with cosmological constant were studied in
\cite{strings}. The issue of black rings turns out to be more
subtle. General considerations do not prevent their existence both
for positive and negative cosmological constants: the additional
centripetal/centrifugal force acting on the $S^1$ can be balanced by
tuning the angular momentum along the $S^1$. And indeed, Chu and Dai
\cite{Chu:2006pf} have found analytically asymptotically $dS$ black
rings within the  $N=4$ de Sitter supergravity (see also
\cite{Gutowski:2010qv}). As in the asymptotically flat (AF) case the
asymptotically de Sitter black holes/rings may have horizon
topologies $S^3$ (or a quotient), $S^1\times S^2$ and $T^3$. For a
negative cosmological constant other topologies can be anticipated,
namely  $S^1\times H^2$ where $H^2$ stands for a negative curvature
hyperbolic two-surface. Meantime, no analytical solutions are known
for black rings with negative cosmological constant. Approximate
``thin'' black rings were obtained by Caldarelli, Emparan and
Rodriguez \cite{CER} using the ``blackfolds'' approach applicable in
arbitrary dimensions \cite{BLF}. These approximate solutions exist
for both signs of the cosmological constant and smoothly go into a
straight black string in the limit of an infinite $S^1$ radius.
Moreover, the possibility of black Saturns with non-flat asymptotics
was also indicated. Supersymmetric black rings with $AdS_5$
asymptotics and compact horizons were investigated in
\cite{Kunduri:2006uh} for a negative cosmological constant.

Lacking exact globally defined solutions, it is tempting to explore
local solutions in the vicinity of the event horizons of the
presumed black objects. This is a particularly fruitful approach in
the extremal case. Usually the near-horizon limits are themselves
exact solutions of the same theory, as in the case of the
near-horizon limit of extremal four-dimensional EM black holes,
which is the Bertotti-Robinson (BR) solution with geometry
$AdS_2\times S^2$ supported by a monopole electric or magnetic
field. Typically, the near-horizon solutions possess a larger
isometry group than the full black hole solutions, therefore they
can be obtained by different solution-generating techniques. All
known exactly five-dimensional supergravity solutions exhibit
enhancement of isometries in the near-horizon region to
$SO(2,1)\times U(1)^2$ (with $U(1)$ factors standing for rotational
symmetries). This is valid for stationary and asymptotically AdS
solutions, and also persists in presence of 
  scalar fields
with a potential \cite{Kunduri:2007vf} and with account for
higher-curvature corrections. In  $D$ dimensions the enhanced
symmetry includes the $U(1)^{D-3}$ rotational symmetry. Note that in
this and more general theories certain properties of extremal black
holes including topology and thermodynamics can be extracted from
the near-horizon solutions using Sen's entropy function approach
\cite{Sen:2005iz}.

Keeping in mind the importance of the BR solution in the
four-dimensional EM theory we explore here similar exact solutions
within the five-dimensional EMCS$\Lambda$ theory with arbitrary
$\gamma$ and $\Lambda$. Presumably these could be near-horizon
limits of black holes/rings with various topologies, including the
above-mentioned case of the hyperbolic topology $S^1 \times H^2$. As
was observed several decades ago, the $n$-dimensional
Einstein-Maxwell theories can be compactified to $(n-2)$-dimensional
spacetime, the two extra space dimensions being curved into a
two-sphere through the action of a monopole magnetic field living on
that two-sphere \cite{HPCS}. This mechanism was later generalized to
the Freund-Rubin compactification of $(d-s)$ or $s$ dimensions by an
$s$-form field \cite{freund}. The consistency of general sphere
compactifications of supergravity actions was extensively discussed
in the past (see, e.g. \cite{Cvetic:2000yp}) showing that a
consistent sphere truncation of gravity coupled to a form field and
(possibly) to a dilaton is possible only in a limited number of
cases.

Our approach consists in compactifying the theory (\ref{emcs5}) on a
constant curvature two-space $\Sigma_2$ of positive, zero or
negative curvature. We consider only compactification through a
direct product ansatz, as in the Freund-Rubin case, thereby avoiding
the consistency problems associated with non-abelian Kaluza-Klein
gauge fields. This compactification, carried out in Sec. 2, results
in the {\em three-dimensional} EMCS$\Lambda$ gravity, with an
additional constraint on the scalar curvature. In Sec. 3 we present
a number of new non-trivial solutions of the five-dimensional
EMCS$\Lambda$ theory of the BR-type which are obtained by uplifting
BTZ, self-dual and G\"odel solutions of the constrained
three-dimensional theory. These solutions are applicable, in
particular, to gauged and ungauged $D=5$ supergravities, but they
are also valid in the EMCS$\Lambda$ theory with more general values
of parameters. In Sec. 4 we perform an alternative toroidal
reduction of the five-dimensional EMCS$\Lambda$ to three dimensions,
deriving a gravity coupled sigma model with a potential. Some of the
solutions listed in Sec. 3 correspond to null geodesics of the
target space. In the case of a vanishing cosmological constant, the
reduced three-space is flat, enabling the generalization of these
solutions to new classes of non-asymptotically flat or
asymptotically flat multi-center solutions of the five-dimensional
theory with arbitrary $\gamma$. The technical proof that the
solutions of Sec. 3 are the only solutions with constant curvature
two-space sections is outlined in the Appendix.

\setcounter{equation}{0}
\section{Reduction on constant curvature two-spaces}
The main idea underlying the present paper is that the
five-dimensional theory (\ref{emcs5}) may be reduced, by monopole
compactification on constant curvature two-surfaces $\Sigma_2$, to
three-dimensional EMCS$\Lambda$:
 \be\lb{ac3}
S_{(3)} = \frac{1}{2\kappa}\int \rd^3x \bigg[\sqrt{|g|} \bigg(\R
-\frac14F^{\alpha\beta}F_{\alpha\beta} - 2\lambda\bigg) -
\frac{\mu}4\epsilon^{\alpha\beta\gamma}F_{\alpha\beta}
A_{\gamma}\bigg]
 \ee
with an additional constraint, which admits several classes of
non-trivial exact stationary solutions, leading to
non-asymptotically flat stationary solutions of the original
five-dimensional theory with the structure $M_3\times \Sigma_2$.

The five-dimensional Maxwell-Chern-Simons and Einstein equations
following from the action (\ref{emcs5}) are
 \ba
\partial_{\mu}(\sqrt{|g\5|}F\5^{\mu\nu}) &=& \frac\gamma{4\sqrt3}
\epsilon^{\nu\rho\sigma\tau\lambda}F_{(5)\rho\sigma}F_{(5)\tau\lambda}\,,
\lb{MCS}\\ {\R\5^{\mu}}_{\nu} - \frac12\R\5\delta^{\mu}_{\nu} &=&
\frac12F\5^{\mu\rho} F_{(5)\nu\rho}-\frac18F\5^2\delta^{\mu}_{\nu} -
\Lambda\delta^{\mu}_{\nu},. \lb{E}
 \ea
In Eq. (\ref{MCS}) we need to fix a sign convention for the
five-dimensional antisymmetric symbol. Throughout this paper we will
assume that $\epsilon^{12345} = +1$, with the spacetime coordinates
numbered according to their order of appearance in the relevant
five-dimensional metric. Let us assume for the five-dimensional
metric and the vector potential the direct product ansatz\"e
 \be\lb{ans2}
\rd s\5^2 = g_{\alpha\beta}(x^{\gamma})\rd x^{\alpha}\rd x^{\beta} +
a^2 \rd\Sigma_k\,, \quad A\5 = A_{\alpha}(x^{\gamma})\rd x^{\alpha}
+ e f_k\rd\varphi\,,
 \ee
where $\alpha,\beta,\gamma=1,2,3$, and the two-metrics for $k=\pm 1,
0$ are
 \ba\lb{Sig}
&&\rd\Sigma_1 = \rd\theta^2 + \sin^2\theta \rd\varphi^2,\quad
\rd\Sigma_0 = \rd\theta^2 +  \theta^2 \rd\varphi^2,
\quad\rd\Sigma_{-1}=\rd\theta^2 + \sinh^2\theta \rd\varphi^2\\
&& f_1 = -\cos\theta,\quad f_0= \frac12 \theta^2,\quad
f_{-1}=\cosh\theta
 \ea
with $\varphi\in [0,2\pi]$ and $\theta\in [0,\pi]$ for $k=1$ and
$\theta\in [0,\infty]$ for $k=0,\,-1$. Here the moduli  $e$ and
$a^2$ are taken to be  constant and real (though for generality we
do not assume outright $a^2$ to be positive). The corresponding five
dimensional geometric quantities are
 \be
\sqrt{|g\5|} = \sqrt{|g|}|a^2|\partial_\theta f_k \,, \quad
R^{\theta}_{(5)\theta} = R^{\varphi}_{(5)\varphi} = k a^{-2}\,,
 \ee
and the Maxwell tensor decomposes as
 \be
F_{(5)\theta\varphi} = e\partial_\theta f_k  \,, \quad
F_{(5)}^{\theta\varphi} = \frac{e}{a^4\partial_\theta f_k}\,, \quad
F=dA\,.
 \ee
Inserting the ansatz\"e (\ref{ans2}) in the field
equations\footnote{Dimensional reduction in the action (\ref{emcs5})
does not produce a correct three-dimensional action.}, we find that
the equations (\ref{MCS}) for $\nu =4,\;5$ are trivially satisfied,
while for $\nu=\beta$ they reduce to
 \be\lb{MCS3}
\partial_{\alpha}(\sqrt{|g|}F^{\alpha\beta}) =
\frac{\gamma
e}{\sqrt3|a^2|}\epsilon^{\beta\gamma\delta}F_{\gamma\delta}\,.
 \ee
The  Einstein equations (\ref{E}) lead to the system
 \ba\lb{E3}
{\R^{\alpha}}_{\beta} - \frac12\R\delta^{\alpha}_{\beta} &=&
\frac12F^{\alpha\gamma} F_{\beta\gamma} -
\frac1{8}F^2\delta^{\alpha}_{\beta} + \left(\frac{4k a^2-e^2}{4a^4}
-
\Lambda\right) \delta^{\alpha}_{\beta}\,, \nn\\
F^2 &=& \frac{4(e^2-3ka^2)}{a^4} + 8\Lambda\,.
 \ea
It is straightforward to check that the reduced equations derive
from the action (\ref{ac3}) with $\kappa=2G_5/|a^2|$ and the
following identification of parameters:
 \be\lb{lamu}
\lambda=\Lambda+(e^2-4k a^2)/4a^4\,, \quad \mu=g/|a^2|\,,\qquad (g =
2\gamma e/\3)\,,
 \ee
with the additional constraint on the three-dimensional scalar
curvature
 \be
\R = (e^2-6ka^2)/2a^4+4\Lambda\,.
 \ee
Inverting the above relations for $k\neq 0$ and $\gamma\neq0$ leads
to
 \be\lb{ae}
a^2 = \frac{k\gamma^2}{3\mu^2/16+(\Lambda-\lambda)\gamma^2}\,, \quad
e =
\frac{\sqrt3\gamma\mu}{2|3\mu^2/16+(\Lambda-\lambda)\gamma^2|}\,,
 \ee
and the constraint
 \be\lb{cons}
\R = \Lambda+3\lambda - 3\mu^2/16\gamma^2 =
\Lambda+3\lambda-e^2/4a^4\,.
 \ee
The equations (\ref{ae}) break down for $k=0$, in which case the
parameters $\mu,\;\lambda$ are related by
 \be
\lambda=\Lambda+\frac{3\mu^2}{16\gamma^2}\,,
 \ee
and for $\gamma=0$, which leads to $\mu=0$.

\setcounter{equation}{0}
\section{Solutions with three commuting Killing vectors}

The three-dimensional theory defined  by the action (\ref{ac3}) is
Maxwell-Chern-Simons electrodynamics (or Maxwell electrodynamics in
the limiting case $\gamma=0$) coupled to cosmological Einstein
gravity. Several classes of exact solutions to this theory with two
commuting Killing vectors and constant Ricci scalar are known
\cite{cam,compere06,tmgebh}. The proof that these are the only
solutions with constant scalar curvature is rather involved and is
given in the Appendix. After uplift to five dimensions according to
(\ref{ans2}) these solutions will lead to Bertotti-Robinson-like
solutions of EMCS$\Lambda$5 with three commuting Killing vectors.

\subsection{BTZ class} The first class corresponds to neutral
(vacuum) three-dimensional solutions with
 \be\lb{curv}
e^2 = 3ka^2 -2\Lambda a^4\,, \qquad \R= 6\lambda = 3\Lambda -
\frac{3k}{2a^2} \,.
 \ee
These exist irrespective of the value of the Chern-Simons coupling
constant $\gamma$. The constant curvature three-space is $dS_3$ for
$\lambda>0$, Minkowski and its coordinate transforms for
$\lambda=0$, and $AdS_3$ and its coordinate transforms, the BTZ
black holes, for $\lambda<0$. We first concentrate on this last
case, before discussing briefly the two other cases $\lambda=0$ and
$\lambda>0$.

The BTZ black hole is a vacuum solution of three-dimensional gravity
with negative $\lambda=-l^{-2}$, which restricts the parameters of
the five-dimensional EMCS$\Lambda$ theory by
 \be
k-2\Lambda a^2>0.
 \ee
This may be considered as a restriction on the five-dimensional
cosmological constant for any given $k$. Combining the BTZ vacuum
black hole with a constant curvature two-surface, we obtain the
following two-parameter family of solutions generically valid for
all $k$:
 \be\lb{bbtz}
\rd s\5^2=-N^2\rd t^2+\frac{\rd r^2}{N^2}+r^2\left(\rd\phi+N^\phi
\rd t\right)^2+a^2 \rd\Sigma_k,
 \ee
where  $\rd\Sigma_k$ is given by(\ref{Sig}), and
 \be
N^2=\frac{r^2}{l^2}-M +\frac{J^2}{4r^2}\,,\quad
N^\phi=\frac{J}{2r^2}.
 \ee
These solutions exist independently of the irrelevant Chern-Simons
coupling constant $\gamma$, including the cases of five-dimensional
minimal supergravity and pure Einstein-Maxwell theory. Generically,
they are supported by magnetic fluxes along $\Sigma_k$, which are
parameterized by $e$ given by
 \be
e^2 = \frac{4a^4}{l^2} + 2ka^2\,,
 \ee
$e=0$ corresponding to $k=-1$ and $\Lambda<0$. The local isometry
group is $SO(2,2)\times SO(3)$ for $k=+1$, $SO(2,2)\times GL(2,R)$
for $k=0$ and $SO(2,2)\times SO(2,1)$ for $k=-1$.

The following particular cases are worth mentioning:

\subsubsection{Minimal supergravity}
In the case $\Lambda=0$, the above relations give, for $a^2>0$,
 \be
k=+1\,, \quad a^2 = \frac{l^2}4 \,, \quad e^2 = \frac{3l^2}4\,.
 \ee
In this case the radius of $AdS$ is twice the radius of the
two-sphere. In the case of minimal supergravity, the solution
(\ref{bbtz}) with $\phi\in R$  coincides with the decoupling
(near-horizon) limit of the general five-dimensional black string
\cite{Compere10}. With $\phi$ periodically identified, the solution
(\ref{bbtz}) may be interpreted as a NAF black ring rotating along
the $S^1$. Moreover, it is the near-horizon limit of the
asymptotically flat black ring with horizon $S^1\times S^2$.

\subsubsection{Gauged supergravity}
The intriguing question about the possible existence of
asymptotically AdS black rings in gauged supergravity is still open.
No supersymmetric black rings are possible in this case, but there
is no proof either of the non-existence of non-BPS rings. Our solution
(\ref{bbtz}) with $\Lambda<0$ and $\phi$ periodically identified
exists in all the three versions $k=1,\;k=0,\; k=-1$, i.e. with
horizon topologies $S^1\times S^2,\; S^1\times R^2,\;$ and
$S^1\times H^2 $. It remains an open question whether these are
indeed the near-horizon limits of topological asymptotically AdS
black rings, but as themselves they can be regarded as NAF and
non-asymptotically AdS rings of various topologies rotating along
$S^1$.

\subsubsection{Rings in De Sitter}
Another yet unsolved issue is the possibility of black rings with
positive cosmological constant, asymptotically dS. As an argument to
support this, we may consider our solutions (\ref{bbtz}) for $k=1$
and $a^2<1/2\Lambda$ as the presumed near-horizon limit of such
rings. In any case, one can interpret these solutions as NAF and
non-asymptotically dS rings of topology $S^1\times S^2,$ rotating
along $S^1$.

\subsubsection{Vacuum solutions}
For $\Lambda<0$, the special case $k=-1,\;l^2=2a^2=-3/\Lambda$ leads
to a two-parameter $(M,\;J)$ vacuum family of NAF topological black
rings, rotating along $S^1$, with horizon topology $S^1\times H^2$.
We are not aware whether this family of locally $AdS_3\times H^2$
topological solutions of the {\em vacuum} five-dimensional Einstein
equations has been reported elsewhere.

\subsubsection{Case $\lambda=0$}
In this case, the constraints (\ref{curv}) imply $e^2=2ka^2$. For
$k=+1$ ($e^2 = 2a^2 = \Lambda^{-1}$, the resulting five-dimensional
metrics include Minkowski$_3\times S^2$, Rindler$_2\times S^1\times
S^2$, and the metric generated from a special coordinate transform
of three-dimensional Minkowski spacetime \cite{C85,CZ94}:
 \be\lb{speflat3}
\rd s\5^2 = -r\,\rd t^2 + 2\,\rd t\,\rd z + \rd r^2 +
\frac1{2\Lambda}\left(\rd\theta^2 +
\sin^2\theta\,\rd\varphi^2\right)\,.
 \ee
The other possibility $k=0$ ($e=0$, $\Lambda=0$), leads to flat
five-dimensional vacuum solutions, including the $k=0$ partner of
(\ref{speflat3}):
 \be\lb{spevac}
\rd s\5^2 = -r\,\rd t^2 + 2\,\rd t\,\rd z + \rd r^2 + \rd\theta^2 +
\rd\varphi^2\,.
 \ee

\subsubsection{Case $\lambda>0$}
In this case, $k=+1$, $\Lambda>0$, and the five-dimensional metric
is the product $dS_3\times S^2$.

\subsection{Self-dual class}

The second class is that of  the ``self-dual'' solutions of
\cite{cam} and \cite{sd} which asymptote to the extreme ($J=Ml$) BTZ
solution (\ref{bbtz}) for $\lambda<0$ or to the three-dimensional
flat metric of (\ref{speflat3}) for $\lambda=0$. For these
solutions, $F^2=0$ (but $F_{\alpha\beta}\neq0$), and the constant
Ricci scalar has again the BTZ value $\R=6\lambda$. The
corresponding five-dimensional solution is, for $\lambda=-l^{-2}$,
$\gamma\neq0$,
 \ba\lb{bsd}
\rd s\5^2 &=& \frac2l\left[-(r - lM_{\mu}(r)/2)\,\rd t^2 -
lM_{\mu}(r)\,\rd t\,\rd z
+ (r + lM_{\mu}(r)/2)\rd z^2 \right] \nn\\
&& \qquad + \frac{l^2}4\frac{\rd r^2}{r^2}
+ a^2\,\rd\Sigma_k\,, \\
A\5 &=& q\left(\frac{2r}{l}\right)^{-\mu l/2}(\rd t - \rd z) +
ef_k\,\rd\varphi\,, \quad M_{\mu}(r) = M - \frac{q^2\mu l}{4(\mu
l+1)}\left(\frac{2r}{l}\right)^{-\mu l} \nn
 \ea
($\mu l \neq 0\,,\, -1$). These solutions depend on the three
independent parameters $a$ (entering $l$, $\mu$ and $e$), and the
two dimensionless parameters $M$ and $q$. For $q=0$ this reduces to
the extreme ($J^2 = M^2l^2$) five-dimensional BTZ solution
(\ref{bbtz}) after the coordinate transformations $r_{BTZ}^2 =
l(r+Ml/2)$, $t_{BTZ} = \2t$, $\varphi_{BTZ} = \2z/l$. These metrics
have generically five Killing vectors (the obvious $\partial_t$,
$\partial_z$ and the three isometries of $\Sigma_k$), except in the
special case $\mu l = -2$, which is the intersection (\ref{bgod1})
of the self-dual and G\"odel classes, with seven Killing vectors.

In the case $\gamma=\mu=0$ of five-dimensional Einstein-Maxwell
theory, the solution degenerates to the solution (derived from the
solution to three-dimensional EM$\Lambda$ theory given in Eq. (29)
of \cite{EML}) with five-dimensional metric given by (\ref{bsd}),
and
 \be
A\5 = q\ln\left(\frac{r}{r_0}\right)(\rd t - \rd z) + ef_k\,
\rd\varphi\,, \quad M_{0}(r) = q^2\ln\left(\frac{r}{r_0}\right)\,.
 \ee
The definition of the mass function $M_{\mu}(r)$ in (\ref{bsd}) also
breaks down for $\mu l = -1$, in which case it must be replaced by
$M_{\mu}(r) = M + (q^2/2l)r\ln(2r/l)$.

The self-dual solution for the case $\lambda=0$ \cite{cam} leads to
the five-dimensional solution
 \ba\lb{bsd0}
\rd s\5^2 &=& -\left(\alpha + \beta r + \frac{q^2}4\,\e^{-2\mu
r}\right)\rd t^2 + 2\,\rd t\,\rd z + dr^2 +
\frac1{2\Lambda}\left(\rd\theta^2 +
\sin^2\theta\,\rd\varphi^2\right)\,, \nn\\
A\5 &=& q\,\e^{-\mu r}\,\rd t -
\frac1{\sqrt\Lambda}\,\cos\theta\,\rd\varphi \qquad
\left(\mu=\frac{2\gamma}{\sqrt{3\Lambda}}\right)\,.
 \ea
This is
horizonless and geodesically complete for $\gamma>0$.

\subsection{G\"odel class}
The third class, corresponding to so-called three-dimensional
G\"odel black holes (no relation with the five-dimensional G\"odel
black holes), was given in \cite{compere06} and \cite{tmgebh} (in
the case where the Chern-Simons term for gravity is absent). These
solutions are closely related to the warped $AdS_3$ black hole
solutions (\ref{bgod}) of topologically massive gravity
\cite{cam,anninos,adtmg,tmgbh,tmgebh}. Using the notations of
\cite{tmgebh}, the three-dimensional solutions, characterized by a
dimensionless constant $\beta^2 = (1-4\lambda/\mu^2)/2$, have a
constant Ricci scalar $\R = (1-4\beta^2)\mu^2/2$, so that the
constraint (\ref{cons}) implies
 \be\lb{lagod}
\lambda= \Lambda+\frac{\mu^2}2\left(1-\frac{3}{8\gamma^2}\right)\,,
 \ee
 leading to
 \be\lb{b2a}
\beta^2 = \frac{k}{\mu^2a^2}-\frac{2\Lambda}{\mu^2} = (k-2\Lambda
a^2)\frac{a^2}{g^2} \,.
 \ee
Comparing (\ref{lagod}) and (\ref{lamu}), we see that for these
solutions the constant $k$ must be related to the five-dimensional
Chern-Simons coupling constant $\gamma$ by
 \be\lb{kga}
k = (3-4\gamma^2)\frac{e^2}{6a^2}\,,
 \ee
so that (assuming $a^2>0$) $k=+1$ for $\gamma<\3/2$, $k=0$ for
$\gamma=\3/2$ and $k=-1$ for $\gamma>\3/2$. The resulting
five-dimensional solution, derived from Eqs. (3.16) and (3.27) of
\cite{tmgebh} (with $c=-1$ to ensure reality of $A\5$, $\kappa=1/2$,
and the coordinate relabellings $t \to z$, $\varphi \to t$) is
 \ba\lb{bgod}
\rd s\5^2 &=& -\frac{\beta^2}{\beta^2-1}\,(r^2-m^2)\,\rd t^2 +
\frac1{\beta^2-1}[(\beta^2-1)\,\rd z - (r+\Omega(\beta^2-1))\,\rd
t]^2 \nonumber \\
&&+ a^2\left[\frac{a^2}{g^2\beta^2}\frac{\rd r^2}{r^2-m^2} +
\rd\Sigma_k\right] \,, \\ A\5 &=& \sqrt{2} \bigg[(\beta^2-1)\,\rd z
- (r+\Omega
(\beta^2-1))\,\rd t\bigg] + e\,f_k\,\rd\varphi\,. \nn
 \ea
where $m^2$ and $\Omega$ are two real parameters\footnote{These
solutions were obtained in \cite{tmgebh} under the assumption $\mu >
0$, i.e. on account of the second Eq. (\ref{ae}) $\gamma e > 0$.}
(the other parameters $\beta^2$ and $e$ depend on the
compactification scale $a$ through (\ref{b2a}) and (\ref{kga})). In
the following we will take $\Omega=0$; the parameter $\Omega$ can be
restored by the local coordinate transformation $z \to z - \Omega
t$. The isometry group of these metrics is $SO(2,1)\times SO(2)
\times SO(3)$ for $\gamma<\3/2$, $SO(2,1)\times SO(2) \times
GL(2,R)$ for $\gamma=\3/2$, and $SO(2,1)\times SO(2) \times SO(2,1)$
for $\gamma>\3/2$ \cite{tmgebh}. Similarly to the metrics
(\ref{bbtz}) of the BTZ class, they are geodesically complete.

\subsubsection{Bertotti-Robinson string}
The geometry described by the metric (\ref{bgod}) depends on the
range of values of the real parameter $\beta^2$. The solution in the
case of pure Einstein-Maxwell theory ($\gamma=0$, implying $k=+1$),
is obtained by
rescaling the coordinate $z \to \beta^{-1}z$, taking the limit
$\beta^2\to\infty$ with $\beta^2g^2=e^2(1-\Lambda e^2)/2$ fixed, and
gauging away the constant $A_{z}$:
 \ba\lb{brstring}
\rd s\5^2 &=& -(r^2-m^2)\,\rd t^2 + \rd z^2 +
\frac{e^2}2\left(\frac1{1-\Lambda e^2}\frac{\rd r^2}{r^2-m^2} +
\rd\theta^2 + \sin^2\theta\,\rd\varphi^2\right)\,, \nonumber \\ A\5
&=& -\sqrt{2}r\,\rd t - e\cos\theta\,\rd\varphi\,,
 \ea
(the three-dimensional reduced solution was previously given in
\cite{EML}, Eq. (25)). This is the five-dimensional embedding of a
four-dimensional dyonic Bertotti-Robinson solution (with
$F\4^2=4\Lambda$ for the ${T^z}_z$ energy-momentum tensor component
to vanish). The geometry is $AdS_2\times S^2\times S^1$.

\subsubsection{Rotating Bertotti-Robinson }
For $\beta^2>1$, the solution (\ref{bgod}), with a regular horizon,
is a five-dimensional analogue of the four-dimensional rotating
Bertotti-Robinson solution $RBR_-$ of EMDA \cite{bremda}. Similarly
to the case of $RBR_-$, this is not a black hole: the parameter $m$
of this solution can be transformed away by a global  coordinate
transformation, and its three-dimensional mass and momentum,
computed according to the prescriptions of \cite{tmgebh}, vanish.

For $\beta^2=1$, the solution (\ref{bgod}) is replaced by
\cite{tmgebh}
 \ba\lb{bgod1}
\rd s\5^2 &=& -(r^2+\alpha)\,\rd t^2 - 2r\,\rd t\,\rd z +\,
a^2\left(\frac{a^2}{g^2}\frac{\rd r^2}{r^2} + \rd\Sigma_k\right)\,, \nn\\
A\5 &=& -\sqrt{2}\,r\,\rd t + ef_k\,\rd\varphi\,,
 \ea
with $\alpha$ a free parameter.

\subsubsection{$H^2\times \Sigma_2\times R$ }
For $0<\beta^2<1$, the coordinate transformation $t \to
(\ol{g}/m)(1-\beta^2)^{1/2}\psi$, $z \to (1-\beta^2)^{-1/2}t$, $r \to
m\cosh\chi$, with
 \be\lb{defb}
\ol{g}=-a^2/\beta^2g\,,
 \ee
transforms (\ref{bgod}) into
 \ba\lb{bgod2}
\rd s\5^2 &=& - \left(\rd t - \ol{g}\cosh\chi\,\rd\psi\right)^2
+\,\ol{g}^2\beta^2\left(\rd\chi^2 + \sinh^2\chi\,\rd\psi^2\right) +
a^2\rd\Sigma_k\,,\nn\\
A\5 &=& -\sqrt{2(1-\beta^2)}\left(\rd t - \ol{g}\cosh\chi\,\rd\psi\right)
+ e\,f_k\,\rd\varphi\,.
 \ea
This metric is horizonless and geodesically complete provided $\psi$
is an angle (period $2\pi$).

\subsubsection{$R^2\times \Sigma_2\times R$ }
For $\beta^2=0$, the regular solution, derived from (3.22) of \cite{tmgebh}
with appropriate coordinate transformations, is
 \ba\lb{bgod0}
\rd s\5^2 &=& -\left(\rd t + \frac{\mu}2x^2\,\rd\psi\right)^2 + \rd x^2
+ x^2\,\rd\psi^2 + a^2\,\rd\Sigma_k \,, \nn\\
A\5 &=& -\sqrt{2}\left(\rd t + \frac{\mu}2x^2\,\rd\psi\right) +
ef_k\,\rd\varphi\,.
 \ea

\subsubsection{$S^2\times \Sigma_2\times R$ with NUT}
Finally, for $\beta^2<0$, the five-dimensional metric (\ref{bgod}) is
replaced by
 \ba\lb{bgod3t}
\rd s\5^2 &=& -\frac{\ol\beta^2}{\ol\beta^2+1}\,(r^2-m^2)\,\rd t^2 -
\frac1{\ol\beta^2+1}[(\ol\beta^2+1)\,\rd z + r\,\rd t]^2 \nonumber \\
&& + a^2\left[-\frac{a^2}{g^2\ol\beta^2}\frac{\rd r^2}{r^2-m^2} +
\rd\Sigma_k\right]
 \ea
(where $\ol\beta^2\equiv-\beta^2$), with signature $(---++)$ in the
coordinate range $r^2>m^2$. However the Minkowskian signature $(+-+++)$
is recovered in the range $r^2<m^2$, which suggests carrying out the
coordinate transformation $t \to (\ol{g}/m)(1+\ol\beta^2)^{1/2}\psi$, $z \to
(1+\ol\beta^2)^{-1/2}t$, $r \to m\cos\chi$, with $\ol{g}$ given by (\ref{defb}),
leading to
 \ba\lb{bgod3}
\rd s\5^2 &=& - \left(\rd t - \ol{g}\cos\chi\,\rd\psi\right)^2
+\ol{g}^2\ol\beta^2\left(\rd\chi^2 + \sin^2\chi\,\rd\psi^2\right) +
a^2\,\rd\Sigma_k\,,\nn\\
A\5 &=& -\sqrt{2(1+\ol\beta^2)}\left(\rd t - \ol{g}\cos\chi\,\rd\psi\right)
+ e\,f_k\,\rd\varphi\,.
 \ea

\subsubsection{Minimal supergravity}
The corresponding G\"odel solution is (\ref{bgod3}) with
 \be
k=-1\,, \quad \ol\beta^2 = \frac18\,, \quad \ol{g}^2\ol\beta^2 = a^2\,, \quad \ol{g}=g\,.
 \ee
By transforming the coordinates $\theta$ and $\chi$ to $x=\cosh\theta$,
$y=\cos\chi$, this solution may be written in the symmetrical form
 \ba\lb{gsym}
\rd s\5^2 &=& -(\rd t - gy\,\rd\psi)^2 + \frac{g^2}8\left[
\frac{\rd y^2}{1-y^2} + (1-y^2)\rd\psi^2  + \frac{\rd x^2}{x^2-1} +
(x^2-1)\rd\varphi^2 \right] \nn\\ A\5 &=& -\frac32(\rd t - gy\,\rd\psi)
+\frac{\3}2\,gx\,\rd\varphi\,,
 \ea
with $x^2>1$, $y^2<1$.

Let us note that, for $\gamma>\3/2$, Eq. (\ref{kga}) can also be solved by $k=+1$,
$a^2=-\ol{a^2}<0$ (reduction on a timelike two-sphere). In the range $r^2 > m^2$, the
resulting metric for minimal supergravity may be written in the form
(\ref{bgod2}), with
 \be
k=+1\,, \quad \beta^2 = -\frac18\,, \quad \ol{g}^2\beta^2 = -\ol{a}^2\,, \quad \ol{g}=-g\,.
 \ee
This metric, with the unphysical signature $(-----)$, is related to (\ref{bgod3})
by the analytical continuation $\chi\leftrightarrow i\chi$,
$\theta \leftrightarrow i\theta$, $\psi\leftrightarrow -\psi$,
$\varphi\leftrightarrow -\varphi$. The corresponding symmetrical form of this ``antiG\"odel''
solution, obtained by putting $x=\cos\theta$, $y=\cosh\chi$, is again (\ref{gsym}),
but with $g\to-g$, and $x^2<1$, $y^2>1$.
Remarkably, as we shall show in a forthcoming paper \cite{brem5}, the G\"odel
and antiG\"odel solutions, with different spacetime signatures, can also be transformed
into each other by $G_{2(+2)}$ sigma-model transformations.

\setcounter{equation}{0}
\section{Multicenter solutions}

\subsection{Toroidal reduction}

All the solutions of EMCS$\Lambda$5 discussed above admit three
commuting Killing vectors. In this case, beside reduction on a
constant curvature two-surface, one can also carry out toroidal
reduction relative to any two $\partial_a$ ($a=1,2$) of these three
Killing vectors, according to the $GL(2,R)$-covariant Kaluza-Klein
ansatz
 \ba\label{st5}
\rd s_{(5)}^2 &=& \lambda_{ab}(\rd x^a + a_i^a\rd x^i)(\rd x^b +
a_j^b\rd x^j) +
\tau^{-1}h_{ij}\,\rd x^i\rd x^j\,, \\
A_{(5)} &=& \sqrt3(\psi_a \rd x^a + A_i\rd x^i)
 \ea
($i,j=3,4,5$) where $\tau = - {\rm det}\lambda$. The Maxwell and
Kaluza-Klein vector fields are then dualized to scalar potentials
$\nu$ (magnetic\footnote{The magnetic potential $\mu$ of
\cite{g2,5to3} is noted here $\nu$ to avoid confusion with the
Chern-Simons coupling constant.}) and $\omega_a$ (twist). In
performing this dualization, we must take care that the scalar
potential $\tau$ can be positive (for most of the solutions
considered here) or negative (in the special case of the G\"odel
solutions (\ref{bgod}) with $\beta^2<0$ and $(5-)$ signature). In
this case $\sqrt{|g\5|} = \varepsilon\tau\sqrt{h}$, where
$\varepsilon =$ sign$(\tau)$, and the dualization equations of
\cite{g2,5to3} are modified to
 \be \lb{dualmu} F^{ij} = a^{aj}
\partial^i \psi_a - a^{ai} \partial^j \psi_a + \varepsilon\frac1{\tau \sqrt{h}}
\epsilon^{ijk} \eta_k\,, \qquad \eta_k =
\partial_k \nu +   \gamma\epsilon^{ab} \psi_a \partial_k\psi_b
\end{equation}
and
\begin{equation}\lb{dualom}
\lambda_{ab}G^{bij} =  \varepsilon\frac1{\tau \sqrt{h}}
\epsilon^{ijk} V_{ak}\,, \qquad V_{ak} =  \partial_k\omega_a -
\psi_a\left(3\partial_k\nu + \gamma\epsilon^{bc}
\psi_b\partial_k\psi_c\right)\,,
\end{equation}
with $G^b_{ij} \equiv \partial_ia^b_j - \partial_ja^b_i$. After
dualization, the reduced field equations derive from the
three-dimensional gravity coupled sigma model with a potential
\begin{equation}\lb{sig}
S_3=\int
 \rd^3x\sqrt{h}\left(-R+\frac12{\cal G}_{AB}\frac{\partial\Phi^A}{\partial x^i}
 \frac{\partial\Phi^B}{\partial x^j}h^{ij} + U\right),
\end{equation}
 where  $R$ is the Ricci scalar of the metric $h_{ij}$ (not to be confused with
the three-dimensional Ricci scalar $\R$ of Sec. 2), $ \Phi^A$
($A=1,\cdots,8$) are the eight moduli $\lambda_{ab}$, $\omega_a$,
$\psi_a$, $\nu$. The potential arises owing to the cosmological
constant and depends on the three moduli $\lambda_{ab}$ through the
only determinant variable $\tau$:
\begin{equation}
U=2\Lambda\tau^{-1}\,.
\end{equation}
The metric of the eight-dimensional target space is:
 \ba\lb{tarmet}
\rd S^2 &\equiv& {\cal G}_{AB}\,\rd\Phi^A\rd\Phi^B = \frac12 {\rm
Tr}(\lambda^{-1}\rd\lambda\lambda^{-1}\rd\lambda) +
\frac12\tau^{-2}\rd\tau^2 - \tau^{-1}V^T\lambda^{-1}V \nonumber\\ &&
+ 3\left(\rd\psi^T\lambda^{-1}\rd\psi - \tau^{-1}\eta^2\right) \,,
 \ea
where $\lambda$ is the $2\times2$ matrix of elements $\lambda_{ab}$,
and $\psi$, $V$ the column matrices of elements $\psi_a$, $V_{a}$.
In the case $\gamma =1$ the target space is a symmetric space with
$G_{2(2)}$ isometry group, while for general $\gamma$ (including
$\gamma=0$) it is not a symmetric space and the isometry group is
solvable.

In the case of the solutions given in the preceding section, the
moduli $\Phi^A$ depend on the three-space coordinates through a
single scalar function which we will denote $\sigma(x)$. The
equations of motion then reduce to
\begin{eqnarray}\label{targeo}
&& \frac{\rD^2\Phi^A}{d\sigma^2}\,\sigma^{,i}\sigma_{,i}
+ \dot\Phi^A\,\nabla^2\sigma = {\cal G}^{AB}\,\partial_B U\,,   \\
&& R_{ij} = \frac12\frac{\rd S^2}{\rd \sigma^2} \,
\sigma_{,i}\sigma_{,j} + U\,h_{ij}\,. \label{qein}
\end{eqnarray}
where $\dot\Phi_A$ stands for the derivative over $\sigma$, $\rD$
denotes the covariant derivative with respect to the target space
metric ${\cal G}_{AB}$, and $\nabla$ is the covariant derivative
with respect to the three-dimensional reduced metric $h_{ij}$. In
the case $\Lambda=0$, the potential $U$ vanishes and, as shown in
\cite{neukra}, the function $\sigma(x)$ can be chosen to be
harmonic,
\begin{equation}\label{harm}
\nabla^2\sigma = 0\,,
\end{equation}
so that Eqs. (\ref{targeo}) reduce to the equations for the target
space geodesics. Null geodesics lead to a Ricci-flat, hence flat,
reduced three-space of metric $h_{ij}$ \cite{spat,bps}. In that
case, the Laplacian $\nabla^2$ becomes a linear operator, so that an
arbitrary number of harmonic functions may be superposed, leading to
a multicenter solution
 \be\lb{sigmult}
\sigma(\vec{x}) = \epsilon +
\sum_i\frac{a_i}{|\vec{x}-\vec{x_i}|}\,.
 \ee
(in the case where the flat three-space is $R^3$).

In what follows we restrict to the case $\Lambda=0$. In the present
case, the target space metric (\ref{tarmet}) does not depend on the
three cyclic coordinates $\omega_a$ and $\nu$ arising from
dualization, so that the corresponding conjugate momenta $\Pi^a$ and
$3 \cal P$ are constants of the motion:
 \be
\Pi^a = - \frac2{\tau}\lambda^{ab}V_b\,, \qquad {\cal P} = - \frac2{\tau}
\eta - \Pi^a\psi_a\,,
 \ee
where we have defined
\begin{equation}
\eta =\dot{\nu}+\gamma\psi ^TJ\dot{\psi}\,,\qquad V=\dot{\omega}
-\left( 3\eta - 2\gamma\psi ^TJ\dot{\psi}\right) \,\psi\,,
\end{equation}
where $\dot{} \equiv \rd/\rd\sigma$, and $J$ is the $2\times2$
matrix $\epsilon_{ab}$. The five remaing geodesic equations and the
null geodesic condition $\rd S^2 = 0$ read
 \be
6\left(\lambda^{-1}\dot{\psi}\right)\dot{} -
\gamma(4\Pi^{\mathrm{T}}\psi
+6\mathcal{P})(J \dot{\psi})-2\gamma(\Pi
^{\mathrm{T}}\dot\psi )(J \psi )
- \left(\frac{3\tau}2(\Pi^{\mathrm{T}}\psi + \mathcal{P}) +
2\gamma\psi ^{\mathrm{T}}J \dot{\psi}\right)\Pi=0\,, \label{psivar}
\ee%
\begin{equation}
\dot{\chi} +\mathrm{Tr}\dot{\chi} + 3 \lambda^{-1}\dot\psi\dot\psi^T
=\frac{\tau
}{4}\left[\Pi\Pi^T\lambda + \Pi ^{\mathrm{T}}\lambda \Pi +3(\Pi ^{\mathrm{T}}\psi
+ \mathcal{P})^{2}\right] \,,  \label{lambdavar}
\end{equation}
\begin{equation}
\frac{1}{2}\mathrm{Tr}(\chi ^{2})+\frac{1}{2}(\mathrm{Tr}\chi
)^{2}-\frac{ \tau }{4}\left[ \Pi ^{\mathrm{T}}\lambda \Pi +3(\Pi
^{\mathrm{T}}\psi + \mathcal{P})^{2}\right]
+3\dot{\psi}^{\mathrm{T}}\lambda ^{-1} \dot{\psi} = 0\,,
\label{null}
\end{equation}
where $\chi \equiv \lambda^{-1}\dot\lambda$.

While it seems difficult to systematically solve this system of
equations, it is easy to promote the special solutions presented in
Sect. 2 to multicenter (null geodesic) solutions, provided that,
after toroidal reduction relative to $\partial_t$ and
$\partial_{z}$, the reduced metric $\rd\hat{s}^2 = h_{ij}\,\rd
x^i\rd x^j$ is flat, with the three possibilities $\rd\hat{s}^2 =$
$\rd r^2 + r^2\rd\Sigma_1$, $\pm\rd r^2 + \rd\Sigma_0$, or $-\rd r^2
+ r^2\rd\Sigma_{-1}$. For $\Lambda=0$, this is the case for the
self-dual solution (\ref{bsd}) (with $k=+1$, $l = 2a$), as well as
its $q=0$ limit, the extreme BTZ solution (\ref{bbtz}) with $J^2 =
M^2l^2$, for the special vacuum solution (\ref{spevac}), for the
G\"odel solution (\ref{bgod1}) with $\beta^2=1$ ($\gamma=1/2$,
$k=+1$, $g^2=a^2$), and for the extreme ($m^2=0$) G\"odel solutions
(\ref{brstring}), (\ref{bgod}) (with $k=+1$, $g^2\beta^2=a^2$) and
(\ref{bgod3t}) (with $k=-1$, $\ol{g}^2\ol\beta^2=a^2$).

\subsection{BTZ and self-dual solutions}
We first consider, in the case $\Lambda=0$, the self-dual solution
(\ref{bsd}) with $k=+1$, $l = 2a$, which contains for $q=0$ the
extreme BTZ solution. In the case of a generic
Chern-Simons coupling constant $\gamma$, the solution (\ref{bsd})
can be generalized by replacing the harmonic function $a/r$ by an
arbitrary harmonic function $\sigma(\vec{x})$,
 \ba\lb{multisd}
\rd s\5^2 &=& \sigma^{-1}\,\rd u\,\rd v + \left(M -
\frac{3c^2\gamma}{4\gamma\pm1}\, \sigma^{\pm4\gamma} \right)
\rd u^2 + \sigma^2\,\rd\vec{x}^2\,, \nn\\
A\5 &=& \3\left[c\,\sigma^{\pm2\gamma}\,\rd u \pm \,A_3\right]
\qquad (\nabla\wedge A_3 = \nabla\sigma)\,,
 \ea
with $u = z-t$, $v = z+t$, $c=q/\3$ \footnote{We have changed a sign
in $A_{(5)}$ because our coordinate transformation implies
$\epsilon_{uv}=-\epsilon_{tz}$.}. The linear superposition
(\ref{sigmult}) leads to multicenter solutions of EMCS$\Lambda$5,
which are asymptotic to the one-center solution (\ref{bsd}) for
$\epsilon=0$, and asymptotically Minkowskian (up to a gauge
transformation) for $\epsilon=1$. These are to our knowledge the
first multi-string solutions of EMCS$\Lambda$5 (multi-hole solutions
were considered in \cite{mikt}).

The one-center asymptotically flat solution may be written in the
ADM form:
 \ba\lb{sdaf}
\rd s\5^2 &=& - \frac{r^2}{(r+a)^2R^2}\,\rd t^2 + R^2 \left(\rd z +
\frac{r}{(r+a)R^2}\,\rd t\right)^2 + \frac{(r+a)^2}{r^2}\rd
r^2 + (r+a)^2\,\rd\Omega_2^2\,, \\
A\5 &=& \3\left[c\left(\frac{r}{r+a}\right)^{\pm2\gamma}(\,\rd z -
\rd t) \pm a\cos\theta\,\rd\varphi\right]\;\; \left(R^2 =  M +
\frac{r}{r+a} - \frac{3c^2\gamma}{4\gamma\pm1}
\left(\frac{r}{r+a}\right)^{\mp4\gamma}\right)\,. \nn
 \ea
For the lower sign the metric has a double horizon at $r=0$. However
this horizon is generically not regular. The first integral for
geodesic motion in the metric (\ref{sdaf}) reads
 \be
\dot{r}^2 + U(r) = \Pi_v^2M
 \ee
with the effective potential
 \be
U(r) = \Pi_u\Pi_v\frac{r}{r+a} +
\Pi_v^2\frac{3c^2\gamma}{4\gamma\pm1}
\left(\frac{r}{r+a}\right)^{\mp4\gamma} + \frac{L^2r^2}{(r+a)^4} -
\frac{\varepsilon r^2}{(r+a)^2}\,,
 \ee
where $\Pi_u$, $\Pi_v$ are the constant momenta conjugate to the
cyclic coordinates $u$ and $v$, $L$ is the orbital angular momentum,
and $\varepsilon = -1$, $0$, or $+1$ for timelike, null, or
spacelike geodesics. It is clear that for the lower sign geodesics
can be analytically continued inside the horizon $r=0$ only for
integer values of $4\gamma$. So (taking into account the fact that
the form of the solution (\ref{sdaf}) breaks down for $\gamma=0$ and
$\gamma=1/4$), the AF solution (\ref{sdaf}) is an extreme black
string for the down sign and $\gamma=(n+2)/4$, $n$ integer (or all
real $\gamma$ for $c=0$). For the upper sign, geodesics with
$\Pi_v\neq0$ are reflected by an infinite potential barrier, while
spacelike geodesics with $\Pi_v=0$ are either reflected or attain
$r=0$ only asymptotically, so that the spacetime is geodesically
complete.

For $\gamma=1/4$ and the lower sign the solution (\ref{multisd}) is
replaced by
 \ba\lb{multisd14}
\rd s\5^2 &=& \sigma^{-1}\,\rd u\,\rd v + \left(M + \frac{3c^2}{4}\,
\sigma^{-1}\ln\sigma \right)
\rd u^2 + \sigma^2\,\rd\vec{x}^2\,, \nn\\
A\5 &=& \3\left[c\,\sigma^{-1/2}\,\rd u - \,A_3\right] \,,
 \ea
and for $\gamma=0$ (Einstein-Maxwell theory) it is replaced by
 \ba\lb{multisd0}
\rd s\5^2 &=& \sigma^{-1}\,\rd u\,\rd v + \left(M - 3c^2\ln\sigma
\right)\rd u^2 + \sigma^2\,\rd\vec{x}^2\,, \nn\\
A\5 &=& \3\left[c\ln\sigma\,\rd u \pm \,A_3\right]\,.
 \ea

In the case of the special vacuum solution (\ref{spevac}), $r$ is
one of the cartesian coordinates of the reduced three-space and is a
harmonic function on that space, so that the generalisation to a
multicenter solution is the vacuum solution \cite{Gib82,spat}
 \be
\rd s\5^2 = 2\,\rd u\,\rd v - \sigma\,\rd u^2  + \rd\vec{x}^2
 \ee
(with $u=t$, $v=z$).

\subsection{G\"odel solutions}
The first obvious candidate multicenter solution in the G\"odel
class is the $\gamma=1/2$, $\beta^2=1$ solution (\ref{bgod1}) with
$k=+1$, $g^2=a^2$. Actually, it turns out that, after a trivial
coordinate transformation $z \propto u$, $t \propto v$, this is just
an instance of the self-dual solution (\ref{multisd}) for
$\gamma=1/2$ and the lower sign, with $M \propto \alpha$.

The Einstein-Maxwell solution (\ref{brstring}) ($\gamma=0$) with
$m^2=0$, $\Lambda=0$ leads to the multicenter solution
 \ba\lb{multibr}
\rd s\5^2 &=& -\sigma^{-2}\,\rd t^2 + \rd z^2 + \sigma^2\,\rd\vec{x}^2\,, \nn\\
A\5 &=& -\sqrt2\left[\sigma^{-1}\,\rd t + A_3\right] \,,
 \ea
which is the trivial five-dimensional embedding of a dyonic
Majumdar-Papapetrou solution.

From the generic $m^2=0$ G\"odel solution (\ref{bgod}) for
$0<\gamma<\3/2$ with $k=+1$, $g^2\beta^2=a^2$, we derive the
multicenter solution
 \ba\lb{multigod1}
\rd s\5^2 &=& -\left(\sigma^{-1}\,\rd t + \rd z\right)^2  +
\beta^2\rd z^2 + \sigma^2\,\rd\vec{x}^2\,, \nn\\
A\5 &=& -\sqrt2\,\sigma^{-1}\,\rd t - \frac{\3}{2\gamma\beta} A_3
\qquad \left(\beta^2 = \frac3{8\gamma^2} - \frac12\right)\,,
 \ea
which may be viewed as a deformation of (\ref{multibr}). For
$\beta^2>1$ ($\gamma<1/2$), the one-center asymptotically flat
solution, generalized by the local coordinate transformation $t \to
t - \varpi z$ (with $\varpi$ a second parameter)
 \ba
\rd s\5^2 &=& -\frac{\beta^2 r^2}{(r+a)^2R^2}\rd t^2 + R^2\left(\rd
z - N^z\,\rd t\right)^2 + \frac{(r+a)^2}{r^2}\,\rd
r^2 + (r+a)^2\,\rd\Omega_2^2\,, \nn\\
A\5 &=& a\left[\frac{\sqrt2}{r+a}\,(\rd t - \varpi\,\rd z) -
\frac{\3}{2\gamma\beta} \cos\theta\,\rd\varphi\right]\,,
 \ea
with
$$R^2 = \beta^2 - (1-\varpi)^2 - \frac{2a\varpi(1-\varpi)}{r+a} +
\frac{a^2\varpi^2}{(r+a)^2}\,, \quad N^z = \frac{r[(1-\varpi)r +
a]}{(r+a)^2R^2}\,, $$ is a dyonic extreme black string.

Finally, for $\gamma>\3/2$,
 \ba\lb{multigod2}
\rd s\5^2 &=& -\left(\sigma^{-1}\,\rd t + \rd z\right)^2  -
\ol\beta^2\rd z^2 + \sigma^2\,\rd\vec{x}^2\,, \nn\\
A\5 &=& -\sqrt2\,\sigma^{-1}\,\rd t - \frac{\3}{2\gamma\ol\beta} A_3
\qquad \left(\ol\beta^2 = \frac12 - \frac3{8\gamma^2}\right)\,,
 \ea
where $\sigma(\vec{x})$ is harmonic on the Minkowskian reduced
metric $\rd\vec{x}^2 = \eta_{ij}\,\rd x^i\,\rd x^j$. The signature
of the metric (\ref{multigod2}) is $(---++)$. A Minkowskian
multicenter solution may be obtained from the $\Lambda=0$ G\"odel
solution (\ref{bgod3}) with $k=-1$, $\ol{g}^2\ol\beta^2=a^2$. By
transforming the $H^2$ coordinates ($\theta$, $\varphi$) to
coordinates ($r$, $z$) such that
 \be
\rd \Sigma_{-1} = \frac{\rd r^2}{r^2} + r^2\,\rd z^2\,, \quad f_{-1}
= r\,,
 \ee
we obtain the multicenter solution with NUTs:
 \ba\lb{multigod3}
\rd s\5^2 &=& -\left(\,\rd t - \ol\beta^{-1}\,A_3\right)^2  +
\sigma^{-2}\rd z^2 + \sigma^2\,\rd\vec{x}^2\,, \nn\\
A\5 &=& -\ds\frac{\sqrt{2(1+\ol\beta^2)}}{\ol\beta}\,A_3 -
\frac{\3}{2\gamma\ol\beta}\sigma^{-1}\,\rd z\,.
 \ea

\section{Summary}

In this paper we have investigated solutions of the general
five-dimensional EMCS$\Lambda$ theory ---containing as particular
cases minimal ungauged and gauged supergravities--- using known
exact solutions of three-dimensional Einstein-Maxwell gravity with a
Chern-Simons term. Our main tool was dimensional reduction, assuming
the five-dimensional spacetime to be the direct product of a
constant curvature surface $\Sigma_2$ of positive, zero or negative
curvature ($k=1$, $0$ or $-1$) and a three-dimensional spacetime.
The reduced theory is then the three-dimensional EMCS$\Lambda 3$
with a constraint on the scalar curvature, which can be satisfied by
three classes of solutions found earlier, namely, the BTZ, self-dual
and G\"odel classes. Promoting these to five dimensions, we have
constructed new EMCS$\Lambda 5$ solutions of the generalized
Bertotti-Robinson type which could be near-horizon limits of black
string and black ring solutions.

The three-parameter BTZ class of solutions is geodesically complete
and exists for an arbitrary (irrelevant) Chern-Simons coupling
$\gamma$. Generically these solutions are non-vacuum, being
supported by magnetic fluxes along the constant curvature
two-surfaces. Their isometry groups are $SO(2,2)\times SO(3)$ for
$k=+1$, $SO(2,2)\times GL(2,R)$ for $k=0$ and $SO(2,2)\times
SO(2,1)$ for $k=-1$. In the minimal supergravity case
$\Lambda=0,\,\gamma=1$ only spherical sections are possible, and the
corresponding solutions are near-horizon limits of black strings.
For a negative cosmological constant this class of solutions exists
in all the three versions $k=1$, $0$, or $-1$, with horizon
topologies $S^1\times S^2$, $S^1\times R^2$ and $S^1\times H^2$
respectively. The latter two could be near-horizon limits of
topological asymptotically AdS black rings, though the existence of
the global solutions remains to be checked. There is a special case
with zero magnetic flux, when we get a two-parameter family of
vacuum solutions which can be seen as non-asymptotically flat
topological black rings with the horizon $S^1\times H^2$. These are
(presumably new) locally $AdS_3\times H^2$ solutions of vacuum
five-dimensional Einstein equations. For a positive cosmological
constant the solutions exists in the $k=1$ version only. Their
interpretation is similar, but the relevant asymptotics is De
Sitter.

The second class of EMCS$\Lambda$5 solutions, depending also on
three parameters, generalize the extremal solutions of the BTZ
class, which they asymptote. They are supported by a magnetic flux
together with an independent dyonic electromagnetic field. The
solutions of the third three-parameter class are generated from
three-dimensional G\"odel black holes. They exist with $k=+1$ for
$\gamma<\3/2$, $k=0$ for $\gamma=\3/2$ and $k=-1$ for $\gamma>\3/2$.
The isometry groups are $SO(2,1)\times SO(2) \times SO(3)$,
$SO(2,1)\times SO(2) \times GL(2,R)$, and $SO(2,1)\times SO(2)
\times SO(2,1)$ respectively. Subclasses include solutions analogue
to previously found rotating BR solutions in dilaton-axion gravity,
horizonless and geodesically complete solutions with spacetime
topology $H^2\times \Sigma_2\times R$, and NUTty solutions with
spacetime topology $S^2\times \Sigma_2\times R$. In the case of
minimal $D=5$ supergravity, the NUTty $S^2\times H^2\times R$
solution (\ref{gsym}) has been shown \cite{brem5} to be a
near-extreme, near-bolt limit of an asymptotically flat solitonic
string solution of EMCS$\Lambda$5.

We have then shown that some of the BR solutions thus constructed
can be promoted to NAF or AF multicenter solutions. For this purpose
we have performed an alternative toroidal compactification of
EMCS$\Lambda$5 to three dimensions, leading to a sigma-model
representation with a potential. For $\Lambda =0$ this potential
vanishes, in which case null geodesics of the target space give rise
to exact solutions of the five-dimensional equations with a flat
reduced three-space. The corresponding BR solutions may be promoted
to multicenter solutions by redefinition of the associated harmonic
functions. We have thus identified three families of multi-string
solutions of EMCS$\Lambda$5, the two ``self-dual'' families
(\ref{multisd}), and the G\"odel family (\ref{multigod1}) (for
$\gamma<\3/2$) or (\ref{multigod3}) (for $\gamma>\3/2$). An
unexpected by-product of our analysis is the construction of new
closed-form asymptotically flat solutions (\ref{sdaf}) generated by
a dyonic electromagnetic field along with a magnetic flux. For the
lower sign these are regular extreme black strings for a discrete
set of values of the Chern-Simons coupling constant (including the
minimal supergravity case $\gamma=1$), while for the upper sign they
are geodesically complete.

\section*{Acknowledgments}
Three of the authors are grateful to the LAPTh Annecy for
hospitality in July (C.-M.C.) and December (A.B. and D.G) 2012 while
the paper was in progress. The work of D.G. was supported in part by
the RFBR grant 11-02-01371-a. The work of C-M.C. was supported by
the National Science Council of the R.O.C. under the grant NSC
102-2112-M-008 -015 -MY3, and in part by the National Center of
Theoretical Sciences (NCTS). The work of A.B. was supported by  the
"Agence Th\'ematique de Recherche  en Sciences et Technologie
(ATRST)" under contract number U23/Av58 (PNR 8/u23/2723) and in part
by the CNEPRU.

\renewcommand{\theequation}{A.\arabic{equation}}
\setcounter{equation}{0}
\section*{Appendix: Constant scalar curvature solutions to three-dimensional EMCS$\Lambda$ theory}

The ansatz \cite{tmgebh}
 \be \lb{par} \rd s^2=\lambda_{ab}(\rho)\,\rd x^a \rd x^b +
\mu^{-2}(\rho)R^{-2}(\rho) \,\rd\rho^2\,, \qquad A = \psi_a(\rho)
\,\rd x^a
 \ee
where $\lambda$ is the $2 \times 2$ matrix
 \be \lambda = \left(
\begin{array}{cc}
T+X & Y \\
Y & T-X
\end{array}\right),
 \ee
and $R^2 \equiv \X^2$ is the Minkowski pseudo-norm of the ``vector''
$\X(\rho) = (T,\,X,\,Y)$,
 \be
\X^2 = \eta_{ij}\,X^iX^j = -T^2+X^2+Y^2 \,,
 \ee
reduces the equations of three-dimensional EMCS$\Lambda$ theory to
the Maxwell-Chern-Simons, Einstein and Hamiltonian constraint
equations
 \ba
&&\S'  = \ds\frac{2}{R^2} \X\wedge\S\,, \lb{S} \\
&& \X'' = \ds\frac{2}{R^2} \bigg[\ds\frac{2}{R^2}(\S \cdot
\X)\X - \S\bigg]\,, \lb{X} \\
&& \X^{'2} + 2\X \cdot \X'' + 4\frac{\lambda}{\mu^2} = 0\,, \lb{ham}
 \ea
where $\S$ is the null vector
 \be
\S = \frac14\left(\psi_0^2 + \psi_1^2\,, \; \psi_0^2 - \psi_1^2\,,
\;  2\psi_0\psi_1\right)\,, \qquad (\S^2 = 0)\,,
 \ee
$\dot{} = \partial/\partial\rho$, and the wedge product is defined
by $({\bf X} \wedge {\bf Y})^i =$ $\eta^{ij}\epsilon_{jkl}X^k Y^l$
(with $\epsilon_{012} = +1$).

To the above equations we must add the constraint (\ref{cons})
 \be\lb{constr}
\R \equiv \mu^2\left[\frac12\X^{'2} - (\X^2)''\right] =
\Lambda+3\lambda - 3\mu^2/16\gamma^2\,.
 \ee
Combining equations (\ref{ham}) and (\ref{constr}), we obtain the
relations
 \be\lb{bc}
(R^2)'' = 2b\,, \quad \X\cdot\X'' = c\,, \quad \X^{'2} = b-c\,,
 \ee
with $b = -(\Lambda+\lambda)/\mu^2 + 3/16\gamma^2$, $c =
(\Lambda-3\lambda)/\mu^2 - 3/16\gamma^2$. From (\ref{X}) it follows
that
 \be\lb{SX}
\S \cdot \X = \frac{c}2R^2\,.
 \ee
Noting that $(\S\cdot\X)' = \S\cdot\X'$ from (\ref{S}), we derive
from (\ref{SX})
 \be\lb{SXsec}
(\S\cdot\X)'' = \S\cdot\X'' + \S'\cdot\X' = c^2 -
\frac2{R^2}\L\cdot\S\,,
 \ee
where we have defined
 \be
\L = \X\wedge\X'\,.
 \ee
Comparing (\ref{SXsec}), (\ref{SX}) and (\ref{bc}), we obtain
 \be\lb{LS}
\L\cdot\S = \frac{c(c-b)}2R^2\,.
 \ee
Next, noting that $\L'\cdot\S = 0$ from (\ref{S}), we obtain
 \be\lb{LS1}
(\L\cdot\S)' = \frac2{R^2}(\X\wedge\X')\cdot(\X\wedge\S) =
\frac2{R^2}\left[(\S\cdot\X)(\X\cdot\X') - R^2(\S\cdot\X')\right] =
-\frac{c}2(R^2)'\,.
 \ee
Comparing with (\ref{LS}), we derive the constraint
 \be\lb{mast}
c(c-b+1)(R^2)' = 0\,.
 \ee

The first possibility $c=0$ means that $\S\cdot\X=0$, which is
equivalent to the self-duality condition \cite{sd}
$F^{\alpha\beta}F_{\alpha\beta}=0$, which leads either to the
self-dual solutions (\ref{bsd}) for $\lambda<0$ and (\ref{bsd0}) for
$\lambda=0$, or to the vacuum solutions of the BTZ class.

Consider now the second possibility, $b-c=1$. Squaring Eq.
(\ref{X}), we find that
 \be\lb{X2}
\X''^2=0\,,
 \ee
while the last equation (\ref{bc}), which now reads $\X^{'2}=1$,
implies that $\X'\cdot\X''=0$. The fact that the null vector $\X''$
is orthogonal to $X'$ implies that the wedge product $\X'\wedge\X''$
is collinear with $\X''$:
 \be\lb{XXX}
\X'\wedge\X'' = \pm \X''\,.
 \ee
This relation, together with Eq. (\ref{X}), may be used to transform
Eq. (\ref{S}) to
 \ba\lb{S1}
\S' &=& - \X\wedge\X'' = \mp \X\wedge(\X'\wedge\X'') \nn\\
&=& \mp(\X\cdot\X')\X'' \pm(\X\cdot\X'')\X' = \mp\frac12(R^2)'\X''
\pm c\X'\,.
 \ea
Taking the scalar product of (\ref{S1}) with the vector $\L$ and
using (\ref{X}) and (\ref{LS}), we obtain
 \be
\L\cdot\S' = \mp\frac{c}2(R^2)'\,,
 \ee
which upon comparison with (\ref{LS1}) shows that we should take the
upper sign in the preceding equations. Finally, Eq. (\ref{S1}) with
the upper sign may be compared with the direct derivative of Eq.
(\ref{X}),
 \be
\S' = -\frac12(R^2)\X''' - \frac12(R^2)'\X'' + c\X'\,,
 \ee
leading to the conclusion that
 \be
\X''' = 0\,.
 \ee
This equation is integrated by
 \be\lb{quad}
\X = \a\rho^2 + \b\rho + \c\,, \qquad (\a^2=0, \quad \a\wedge\b =
-\a)\,,
 \ee
the vector relations between the constant vectors $\a$, $\b$, $\c$
following from (\ref{X2}) and (\ref{XXX}). We recognize in
(\ref{quad}) the quadratic ansatz which was used in \cite{tmgebh} to
derive the G\"odel solutions to three-dimensional EMCS$\Lambda$.

For the last possibility, $(R^2)'=0$, we can adapt the above
argument to show that
 \be\lb{XXX0}
\X'\wedge\X'' = q\X'' \qquad (q^2=b-c)\,.
 \ee
Eq. (\ref{S1}) is now replaced by
 \be
\S' = \frac{c}q\X'\,.
 \ee
Taking the scalar product with $\X'$ and using (\ref{SXsec}) and
(\ref{LS}), we obtain
 \be
\S'\cdot\X' = cq = cq^2\,.
 \ee
The first solution, $c=0$, leads to a subcase of the self-dual
class. The second solution, $q=0$, means that $\X''$ is collinear
with $\X'$ so that $\X\cdot\X'' = c = 0$, leading again to a subcase
of the self-dual class. And the last solution, $q=1$, means $b-c=1$,
corresponding to a subclass of the G\"odel class.

\end{document}